# The Quantum Abacus: Analog Computing using Surface Rydberg States

Peter B. Lerner[1]

Abstract

Recently, Rydberg atoms appeared as a viable alternative to the quantum gates built on atomic or molecular ions. The lifetimes of the circular Rydberg states can be in the millisecond range. That prevents inherent metastability of the Rydberg atoms to influence computation at the typical decoherence times, which are now being achieved in the range of 1 μs. The paper proposes to use a pinning potential of an image charge on a cryogenic substrate (liquid He, in particular) to confine large densities of Rydberg atoms to dielectric surfaces. The substrate can also act as a natural cooler medium. A design of the computer (quantum abacus) based on these ideas is briefly sketched in the paper.

---

[1] Wenzhou Kean University and SciTech Associates

Quantum computing presents a self-contradictory challenge to the designer. The gates have to be sufficiently isolated to avoid decoherence, yet significantly and controllably coupled to their environment to upload the inputs of a computer program and to download the output with a significant speed.

Current projects of the quantum computer involve several attempts at realization: trapped ions and atoms, including Rydberg states [1, 2]; arrays of Josephson junctions and N-vacancies in diamond-like materials.

Trapped ions and atoms were the first configuration to be practically implemented. [1] It is not surprising despite the exotic nature of the system because ions and atoms in a particle trap are the easiest to isolate from decohering in a random environment. However, the QC built on this elemental base suffers from a scalability problem. Because of a possible chaotic motion of particles in a trap, they have to be individually controlled, which effectively limits possible number of gates to several tens.

Recently, Rydberg atoms appeared as a viable alternative to the gates built on atomic or molecular ions. The lifetimes of circular Rydberg states can be in the millisecond range. [3] That prevents an inherent metastability of the Rydberg atoms to influence computation at decoherence times, which are now being achieved in the range around 1 μs. [4] The breakthrough in the use of Rydberg atoms in computation came in the form of the demonstration of the Coloumb blockade, which allows Rydberg atoms to be individually addressed. [5] Yet, their stochastic motion in a trap presents a significant liability for their use as elemental base for anything reminiscent of a practical computer. In particular, creating large and uniform densities of Rydberg atoms in a "free" space of particle trap is a problem of enormous experimental complexity. This is in part due to the fact that a resonance transfer of energy between Rydberg atoms can easily lead to the ionization because of extremely large ionization and de-excitation cross-sections [6].

Yet, the author, together with I. M Sokolov[2] in the late 1980s proposed to use a pinning potential of an image charge on a cryogenic substrate (liquid He, in particular) to confine Rydberg atoms to dielectric surfaces (P. B. Lerner, I. M. Sokolov, 1986, 1989, [7], [8]). The substrate can also serve as a natural cooler medium, through the phonon coupling to the liquid ripples. We show a schematic drawing of the operational and control levels/bands in Fig. 1. Previously, electronic spectra of the isolated surface-state electrons on liquid $^3$He, $^4$He was studied by Konstantinov, Issiki et al. [9]. The scattering of the surface state electrons (pinned in the direction perpendicular to the surface, unlike the proposed scheme) on the self-excited ripples was studied in [10].

---

[2] Currently, a chair in theoretical physics at Humboldt U., Berlin

One of the experimental possibilities for creating such atoms, which has been proposed by the authors was to excite He atom into a Rydberg state directly from the helium substrate. This experimental possibility was realized by the DESY group (von Haeften, etc.) in 2001 [11] In particular, they clearly resolved inter-subband transitions, which are instrumental for the operation of the proposed quantum computer (see Fig. 1).

One of the problems of the Rydberg-atom based computing is a difficulty of miniaturization. The transitions for Rydberg atoms usually lie in the microwave to far-IR spectral range. Henceforth, the steering antennas for the device must necessarily be large compared to the basic components of the modern electronics. To reduce the size of the steering antennas, we propose to use spoof plasmons propagating in the THz range. (see, e.g. [12] and *op. cit.*)

A resulting device, the quantum abacus is shown in the Fig. 3. It represents a corrugated gold template covered with liquid He. The film of the liquid He covering the template distinguishes this proposal from the experiment by J. Naber et al. [13] Rydberg atoms are being pinned to the substrate by the modulated image forces—stronger in the shallower areas and weaker in the deeper areas of the plate (Fig. 1). Channels in the substrate were first used in the experiments by Sabouret et al. [14] Using surface state electrons being pinned to their atoms in highly excited Rydberg states distinguishes our scheme for the computer from the experiment by Ge Yang et al. [15]

As is shown in Fig. 1 when the atomic levels of the Rydberg atom are excited, the atom receives a recoil momentum from the substrate, which can be potentially exploited as moving an atom to the next "register." Steering waveguides can be positioned below the surface of the metal so the microwave or THz radiation in the waveguides couples to the plasmons at the interface of the corrugated gold plate and the Helium liquid (Fig. 3)

**Figure captions**

Fig.1 Schematic wavevector-energy diagram for the level mixing between plasmon-polaritons and the Rydberg atomic level.

Fig. 2. Pinning of the Rydberg atoms by the image potential—weaker on a deep helium liquid, stronger—in the close presence of the substrate.

Fig. 3. Schematic drawing of the quantum abacus. Rydberg atoms are pinned to the helium liquid, unless they are shifted through excitation (see Fig. 1). Spoof plasmons in the corrugated metallic substrate are excited by EM waves (red arrows) in the waveguides under the substrate.

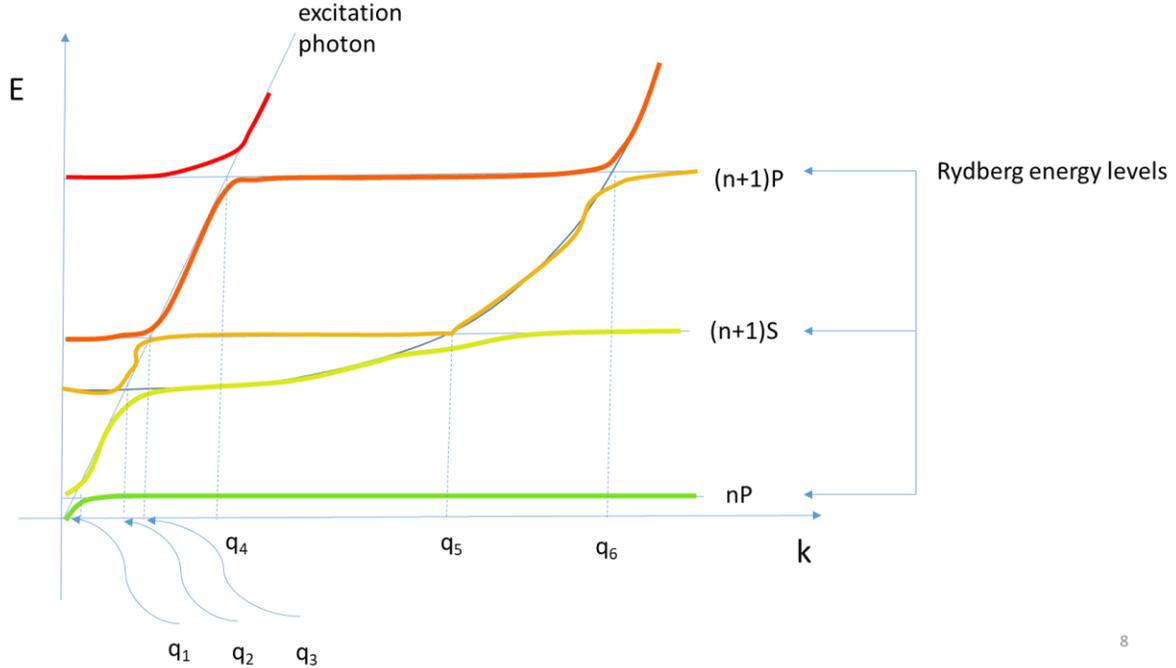

Fig.1

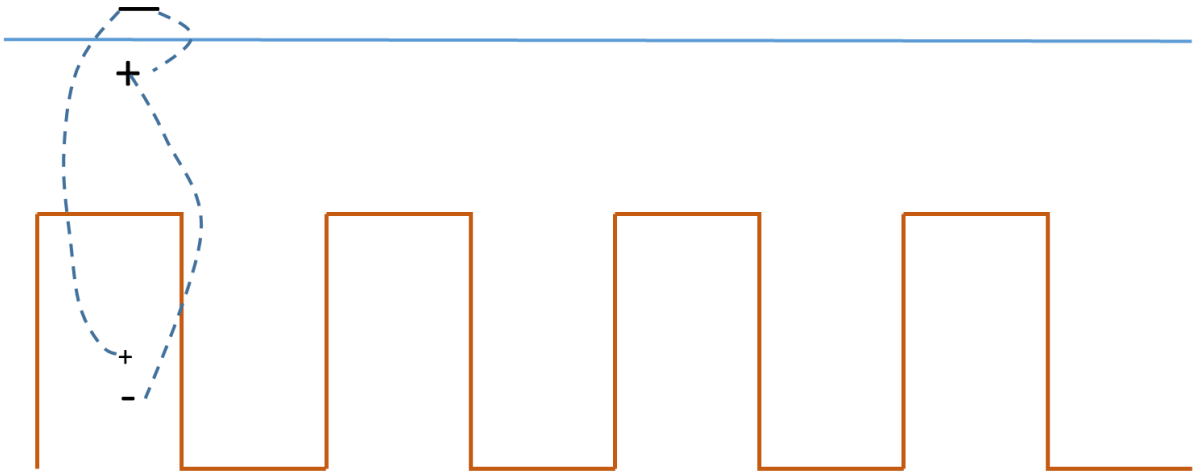

Fig. 2

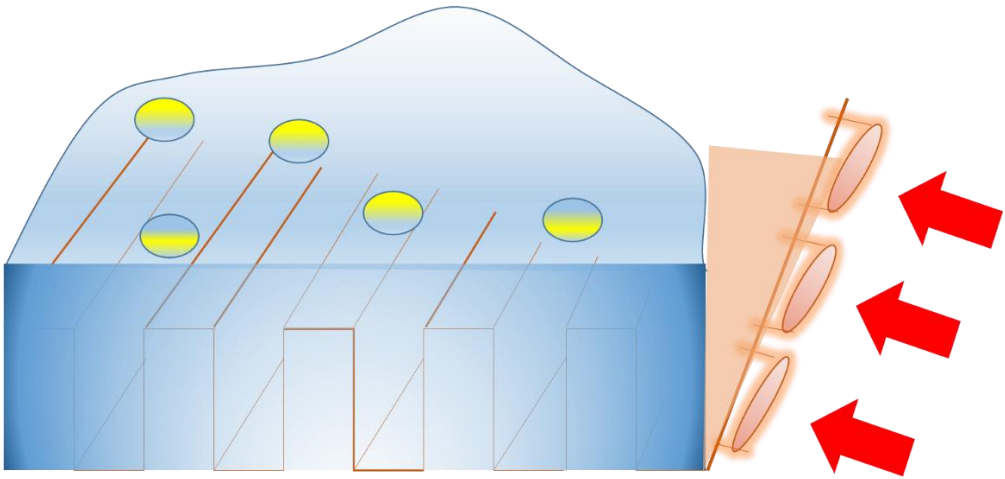

Fig. 3